\documentclass[twocolumn]{svjour3}         
\smartqed  
\usepackage{graphicx}
\usepackage{astron}
\usepackage{amsmath,amssymb,amsfonts} 
\usepackage{color}                    
\usepackage{hyperref}                 
\usepackage{fancyhdr} 
\usepackage{longtable} 

\begin{document}
\title{High angular resolution imaging of the circumstellar material around
intermediate mass (IM) stars
}


\author{A. Fuente}


\institute{A. Fuente \at
Observatorio Astron\'omico Nacional (OAN)\\ 
Apdo 112, E-28800 Alcal\'a de Henares, Spain\\
\email{a.fuente@oan.es}
}

\date{Received: date / Accepted: date}

\maketitle

\begin{abstract}

In this Paper we present high angular resolution
imaging of 3 intermediate-mass (IM) stars 
using the Plateau de Bure 
Interferometer (PdBI). In particular we present
the chemical study we have carried out towards the 
IM hot core NGC~7129--FIRS~2.
This is the first chemical study in an IM hot core and
provides important hints to understand
the dependence of the hot core chemistry on the stellar luminosity. 
We also present our high angular resolution 
(0.3$"$) images of the borderline Class~0-Class~I object IC1396~N. 
These images trace the warm region of this IM protostar
with unprecedent detail (0.3$"$$\sim$200~AU at the distance
of IC1396~N) and provide the first detection of a cluster of
IM hot cores.  Finally, we present our interferometric continuum
and spectroscopic images of the 
disk around the Herbig Be star R~Mon. We have determined
the kinematics and physical structure of the disk associated
with this B0 star. The low spectral index derived from the
dust emission as well as the flat geometry of the disk 
suggest a more rapid evolution of the 
disks associated with massive stars (see Poster by Alonso-Albi et al.).
In the Discussion, we dare to propose a possible evolutionary
sequence for the warm circumstellar material around IM stars.


\end{abstract}

\section{Introduction}
\label{intro}
Luminous intermediate-mass young stellar objects (IMs) (protostars and stars with 
M$_*$$\sim$5--10~M$_\odot$) are crucial in star formation studies.
They share many characteristics with high-mass stars (clustering, PDRs) 
but their study presents an important advantage: there are many 
located closer to the Sun (d$\sim$1~Kpc), and in regions less complex 
than massive star forming regions. Thus, they can be studied
with high spatial resolution. The study of
important problems of massive star formation
like the physical and chemical structure of hot cores,  
clustering and the occurrence and physical
properties of the disks around massive stars, requires of high 
spatial resolution. With the current instrumentation, these problems 
can only be addressed in IMs.

During the last 3 years, we have mapped a sample of IMs
in different evolutionary stages using the Plateau de Bure 
Interferometer (PdBI). In particular we have detected and carried
out a chemical study towards the IM hot core embedded in the 
Class 0 object NGC~7129--FIRS~2. 
Moving to more evolved objects, we have studied in detail
the circumstellar disk associated with the Herbig Be star R Mon
(see Poster by Alonso-Albi et al.). Finally,
we have imaged the borderline
Class~0-Class~I object IC1396~N using the new A configuration of the PdBI
which provides the highest angular resolution that can be achieved with the
current millimeter instrumentation (0.3$"$$\sim$200~AU at the distance
of IC1396~N). These observations have allowed us to detect
a cluster of hot core/corinos and have a first glance at the evolutionary stage of 
each cluster component. In this Paper, we revise the results obtained
from these high spatial resolution studies and discuss 
the implications for the undestanding of the massive star formation process. 


\section{NGC~7129--FIRS~2}
\label{n7129}

NGC~7129--FIRS~2 (hereafter, FIRS~2), with 
a luminosity $\sim$ 500 L$_\odot$ and 
a stellar mass $\sim$5~M$_{\odot}$ , is very likely the
youngest IM object known at present (Fuente et al. 2001,2005a). 
Recent PdBI observations in the continuum and spectroscopic lines
carried out by our team show the existence of an IM hot core towards 
this young protostar (Fuente et al. 2005b). We 
estimate a size of 650$\times$900~AU and a mass
of 2~M$_\odot$ for the hot core associated with this object.
The dimensions and mass of this IM hot core are intermediate between  
those measured in hot corinos
(r$\sim$150~AU,M$<$1~M$_\odot$) and massive stars 
(r$\sim$3000~AU,M$>$10~M$_\odot$) and consequently, a differentiated
chemistry is expected. This IM hot core provides 
a unique opportunity to study the dependence of the hot core
chemistry on the stellar luminosity.

A large number of molecular lines have been detected in our interferometric spectra
towards FIRS~2. Most of these lines are identified as belonging to deuterated
(D$_2$CO, c-C$_3$D and c-C$_3$HD), sulphuretted ($^{13}$CS, OCS), and complex
O-/N-bearing species (HCOOH, C$_2$H$_5$OH, C$_2$H$_5$CN). One chemical
difference between hot corinos and hot cores is the enhanced
abundance of deuterated species in the former.
Loinard et al. (2003) searched for the doubly deuterated form of 
formaldehyde (D$_2$CO) in a large 
sample of young stellar objects. D$_2$CO was detected in all low-mass protostars 
with [D$_2$CO]/[H$_2$CO] 
ratios of 2--40\%. On the other hand, no detection was obtained 
towards more massive protostars, where [D$_2$CO] [H$_2$CO]$<$0.5\%.
We estimate a [D$_2$CO]/[H$_2$CO]$\sim$0.14 towards the IM hot core
FIRS~2. This value is 4 orders of magnitude larger than the cosmic D
abundance and similar to those found in pre-stellar clumps and
low-mass protostars. 
 
The sulphuretted and complex compounds
are characteristic of hot cores in both the low-mass and the high-mass regimes.
We have compared the abundances of complex molecules in FIRS~2 with those
in hot corinos and the massive hot cores OMC1 and G327.3--0.6. Contrary to
model predictions, we did not detect any dependence of the O-/N-complex molecules
ratio on the protostellar luminosity. However, we detected differences between the
behavior of the O-bearing species with the stellar luminosity. While H$_2$CO and
HCOOH are more abundant in low luminosity sources, CH$_3$OH 
seems to be more abundant in massive objects (see Fig. 1). 
Fuente et al. (2005b) proposed that this could be due to a different mantle
composition in the two classes of region, caused by different
physical conditions (mainly gas density and dust temperature)
during the pre-stellar and accretion phase.
However, this could also be
due to other factors, such as the different spatial scale of the observations or
a possible contribution of the shocked gas associated with the bipolar outflow 
to the emission of these molecules. 
The detection
and detailed study of other
intermediate-mass and low mass hot cores are necessary 
to establish firm conclusions.

\setlength\unitlength{1cm}
\begin{figure}
\vspace{12.5cm}
\includegraphics{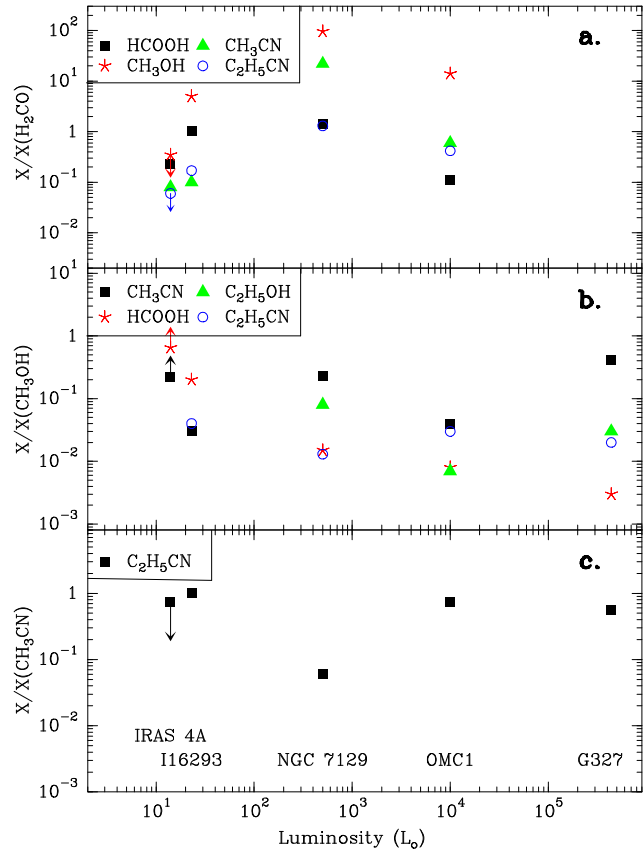}
\caption{Relative abundances of the complex O- and N-bearing molecules
as a function of the protostellar luminosity for a sample of hot cores/corinos.
Note that the HCOOH abundance seems to decrease with the protostellar
luminosity (Fuente et al. 2005b).}
\end{figure}

\setlength\unitlength{1cm}
\begin{figure}
\vspace{6.5cm}
\includegraphics{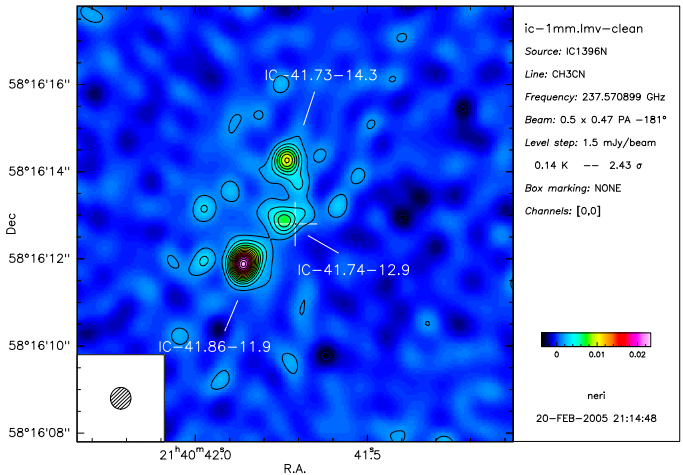}
\caption{Interferometric continuum image at 1.3mm obtained with the Plateau
de Bure Interferometer (PdBI) towards the Class 0 protostar IC~1396N. A cluster
of at least 3 hot cores cores is detected towards BIMA~2 (Neri et al. 2007).}
\end{figure}

\section{R~Mon}
\label{obs}

R~Mon is the most massive disk detected in
dust continuum emission at mm wavelengths 
around a Herbig Be star (Fuente et al. 2003). 
Moreover, it is
the only one detected in 
molecular lines and thus far, our unique
opportunity to investigate the physical structure 
and kinematics of the disks
associated with these stars. 

The high angular resolution continuum images at 2.7mm
and 1.3mm reported by
Fuente et al. (2006) allow us to determine the position
(R.A.=06:39:09.954  Dec=+08:44:09.55 (J2000) and
size ($\sim$150~AU) of the dusty disk.  
Moreover, by fitting the SED at cm and
mm wavelengths we determine a disk mass of 0.007~M$_\odot$ 
and $\beta$=0.3--0.5. 
Values of $\beta$ between 0.5 and 1
are usually found in circumstellar
disks around HAE and TTs and are thought to be
evidence for grain growth in these disks (see e.g.
Natta et al. 2006).
The low value of $\beta$ in R~Mon suggests that
grain growth has proceeded to very large sizes already in the short lifetime
of its disk.

Alonso-Albi et al. report 
interferometric $^{12}$CO
and $^{13}$CO observations towards this
disk. They conclude that the disk is in Keplerian rotation
around the star. Keplerian rotation has also been 
found in most of the TTs and Herbig Ae stars 
studied thus far
and indicates a similar formation mechanisms for
the stars in the range 1--8 M$_\odot$. However,
contrary to the low mass T Tauri and Herbig Ae
stars that are usually surrounded by flared disks,
the observed $^{12}$CO/$^{13}$CO intensity ratio
in R~Mon shows that the disk is geometrically flat 
(see Poster by Alonso-Albi et al.). This result is
in line with previous near-IR and optical measurements that
suggest that 
Herbig Be stars have geometrically flatter disks than
Herbig Ae and T Tauri stars (see e.g. 
Meeus et al. 2001)

The flattening of the disk in Herbig Be stars 
can be due to the rapid grain growth. 
The grain growth causes the optical depth of the
disk to drop and allows the UV radiation to penetrate
deep into the circumstellar disk and photo-evaporate the disk external
layers (Dullemond \& Dominik 2004). Thus, we can propose
an evolutionary sequence in which
the disks associated with Herbig Be stars
start with a flaring shape but become
flat during the pre-main sequence
and lose most of their mass ($>$90\%) before the star
becomes visible ($<$10$^5$ yr).
\begin{table*}
\caption{Millimeter flux densities, sizes, spectral indexes and masses}             
\label{table:1}      
\centering          
\begin{tabular}{l c c c c c }     
\hline\hline       
\multicolumn{6}{c}{3.3mm (91.7 GHz)} \\
\hline                    
\multicolumn{1}{c}{} & \multicolumn{1} {c}{BIMA 3}  &  \multicolumn{4}{c}{BIMA 2} \\
                                 &                       &  Cocoon    &    41.86+11.9   &  41.73+12.8    &  41.73+14.3  \\ \hline
$\alpha$ (J2000)     & 21:40:42.84    &   21:40:41.86  & 21:40:41.85     &    21:40:41.73  & 21:40:41.72  \\
$\delta$  (J2000)     & 58:16:01.4      &   58:16:13.2    & 58:16:11.9       &    58:16:12.8    &  58:16:14.3   \\
Size ($"$)                &   0.8 x 0.5        &    4.3 x 3.1     &   unresolved    &     unresolved  &   unresolved  \\        
S (mJy)                   &         8              &        16           &         5.9           &          1.5          &         2.7  \\ \hline 
\multicolumn{6}{c}{1.3mm (237.6 GHz)}\\
\multicolumn{1}{c}{} & \multicolumn{1} {c}{BIMA 3}  &  \multicolumn{4}{c}{BIMA 2} \\
                                 &                       &  Cocoon    &    41.86+11.9   &  41.73+12.8    &  41.73+14.3  \\ \hline
$\alpha$ (J2000)    &   21:40:42.84    &   21:40:41.85   & 21:40:41.86   &   21:40:41.73     &  21:40:41.73 \\ 
$\delta$ (J2000)     &   58:16:01.4      &   58:16:13.1     & 58:16:11.9     &   58:16:12.8       &  58:16:14.3 \\
Size ($"$)               &     0.8 x 0.4        &   4.5 x 3.1       &   0.4 x 0.2       &   unresolved      &   unresolved \\
S (mJy)                  &        30               &       245           &       35             &          6               &     10      \\ \hline
\multicolumn{1}{c}{} & \multicolumn{1} {c}{BIMA 3}  &  \multicolumn{4}{c}{BIMA 2} \\
                                 &                       &  Cocoon    &    41.86+11.9   &  41.73+12.8    &  41.73+14.3  \\ \hline
Mean Spec.Index   &      1.4               &    \multicolumn{4}{c}{2.6} \\
Spec.Index             &       1.4              &          2.8          &       1.9            &         1.5             &         1.4  \\
Mass (M$_\odot$)   &     0.05             &         0.4                      &       0.06          &       0.01             &     0.01                \\
$\beta$                    &     $\leq$0        &        $\sim$1.0             &                       &                           &                       \\
\hline                  
\end{tabular}
\end{table*}

\section{IC1396~N}
 
The new A configuration of the  PdBI allows us to study the 
warm interior of protostellar envelopes with unprecedent detail.
In this Paper, we present interferometric continuum observations of
the IM protostar IC1396~N (Neri et al. 2007). 

IC1396~N is a 440~L$_\odot$
protostar located at a distance of 
750~pc and Classified as a Class 0/I borderline source. 
Figure 2 shows the continuum image at 1.3mm. 
This image shows the presence of at least three bright continuum
emission sources in the center of IC1396~N, all three
associated with the source identified as BIMA~2  by Beltr\'an et al. (2002). 
The source
BIMA~3 is also detected in our continuum image although lies outside
the region shown in Figure 2. In addition to 
the 3 compact cores we detect some kind of extended emission in
BIMA~2. 
In fact, the bulk of the mm-emission is emerging from a large
region ($\sim$3000~AU) centered on the triple-system. 
According to these results, 
we envision two different models for the continuum emission: 
(a) an envelope with sharp boundaries in which the three 
compact cores are embedded, 
(b) a region harboring a cluster of lower brightness
cores from which we have detected  the three most intense ones. 
The lack of sensitivity to large scale emission and
emission distributed on a large number of cores makes it difficult to
argue against one or the other model. For simplicity, we favor
the model of the dusty `cocoon' in which the three intense cores are embedded.

Neri et al. (2007) modeled the emission at 2.7mm and 1.3mm in the UV-plane
assuming a `cocoon+cores' structure. In Table 1
we show the fluxes and spectral indexes derived
from this model. The weaker cores were not
resolved by the interferometer. The primary core 41.86+13.2 is resolved
in the 1.3mm emission to a size of $\sim300$\,AU$\times 150$\,AU, i.e.
an order of magnitude larger than the size measured in hot corinos.
This is consistent with this source being the precursor of a Herbig Ae/Be star. 
The `cocoon' accounts for the 80\% of the 1.3mm continuum
emission and has a different spectral index from the compact 
cores. While the
spectral index in the extended emission is $\sim$2.8 as expected
for dust continuum emission with a standard value of $\beta$=1,
the spectral indexes of the compact hot cores are all $<$2.
We propose that this
change in the spectral index is very likely associated with a change of the
grain properties. The grains in
the compact hot cores might be  similar to those found in
the evolved cirumstellar disks. In fact, the compact
hot cores could be actually disks. We cannot discard, however,
other possible interpretations for these small compact regions (see Neri et al. 2007
for a more detailed discussion).

\section{Summary and Discussion}
A major interest in the Astrophysics today is the understanding of
hot cores. These warm regions in the interior of
the protostellar envelopes are the prime material from which the proto-planetary disks
are formed. However the detailed physical and chemical structure of
these hot regions as well as their evolution to become proto-planetary
disks are not known. High spatial
resolution imaging of the warm regions of protostellar objects is
required to have a deeper insight into this problem. 

In this Paper we present high angular resolution
imaging of 3 IMs using the Plateau de Bure 
Interferometer (PdBI). These stars are thought to be
in a different evolutionary stage.  
NGC~7129--FIRS~2 is a young Class 0 object hosting a
massive ($\sim$2~M$_\odot$) and compact hot core.
IC1396~N is a borderline Class~0-Class~I object. Our interferometic
images reveal a massive `cocoon' in which three compact cores are embedded.
R~Mon is a visible star surrounded by a circumstellar disk.
We can propose a simple evolutionary scheme in which a IM star
starts its life surrounded by a massive and dense hot core. 
The newly formed star(s) disperses part of the hot core 
material producing a less dense `cocoon' in which the
circumstellar disks are immersed. At this stage, the density
contrast between the `cocoon' and the molecular cloud is still
high enough
for the cocoon to be detected with the interferometer. 
Finally the cocoon is dispersed and only the circumstellar
disk is left.
Of course, this is only
a rough description of what the evolution of the warm 
circumstellar material could be and future interferometric 
observations are needed
to confirm and extend this scheme.

\begin{acknowledgements}
This work has been partially supported by the Spanish MEC and 
Feder funds under grant ESP2003-04957 and by SEPCT/MEC 
under grants AYA2003-07584.
\end{acknowledgements}

\def\jnl@style{\it}
\def\aaref@jnl#1{{\jnl@style#1}}

\def\aaref@jnl#1{{\jnl@style#1}}

\def\aj{\aaref@jnl{AJ}}                   
\def\araa{\aaref@jnl{ARA\&A}}             
\def\apj{\aaref@jnl{ApJ}}                 
\def\apjl{\aaref@jnl{ApJ}}                
\def\apjs{\aaref@jnl{ApJS}}               
\def\ao{\aaref@jnl{Appl.~Opt.}}           
\def\apss{\aaref@jnl{Ap\&SS}}             
\def\aap{\aaref@jnl{A\&A}}                
\def\aapr{\aaref@jnl{A\&A~Rev.}}          
\def\aaps{\aaref@jnl{A\&AS}}              
\def\azh{\aaref@jnl{AZh}}                 
\def\baas{\aaref@jnl{BAAS}}               
\def\jrasc{\aaref@jnl{JRASC}}             
\def\memras{\aaref@jnl{MmRAS}}            
\def\mnras{\aaref@jnl{MNRAS}}             
\def\pra{\aaref@jnl{Phys.~Rev.~A}}        
\def\prb{\aaref@jnl{Phys.~Rev.~B}}        
\def\prc{\aaref@jnl{Phys.~Rev.~C}}        
\def\prd{\aaref@jnl{Phys.~Rev.~D}}        
\def\pre{\aaref@jnl{Phys.~Rev.~E}}        
\def\prl{\aaref@jnl{Phys.~Rev.~Lett.}}    
\def\pasp{\aaref@jnl{PASP}}               
\def\pasj{\aaref@jnl{PASJ}}               
\def\qjras{\aaref@jnl{QJRAS}}             
\def\skytel{\aaref@jnl{S\&T}}             
\def\solphys{\aaref@jnl{Sol.~Phys.}}      
\def\sovast{\aaref@jnl{Soviet~Ast.}}      
\def\ssr{\aaref@jnl{Space~Sci.~Rev.}}     
\def\zap{\aaref@jnl{ZAp}}                 
\def\nat{\aaref@jnl{Nature}}              
\def\iaucirc{\aaref@jnl{IAU~Circ.}}       
\def\aplett{\aaref@jnl{Astrophys.~Lett.}} 
\def\apspr{\aaref@jnl{Astrophys.~Space~Phys.~Res.}}
\def\bain{\aaref@jnl{Bull.~Astron.~Inst.~Netherlands}} 
\def\fcp{\aaref@jnl{Fund.~Cosmic~Phys.}}  
\def\gca{\aaref@jnl{Geochim.~Cosmochim.~Acta}}   
\def\grl{\aaref@jnl{Geophys.~Res.~Lett.}} 
\def\jcp{\aaref@jnl{J.~Chem.~Phys.}}      
\def\jgr{\aaref@jnl{J.~Geophys.~Res.}}    
\def\jqsrt{\aaref@jnl{J.~Quant.~Spec.~Radiat.~Transf.}}
\def\memsai{\aaref@jnl{Mem.~Soc.~Astron.~Italiana}}
\def\nphysa{\aaref@jnl{Nucl.~Phys.~A}}   
\def\physrep{\aaref@jnl{Phys.~Rep.}}   
\def\physscr{\aaref@jnl{Phys.~Scr}}   
\def\planss{\aaref@jnl{Planet.~Space~Sci.}}   
\def\procspie{\aaref@jnl{Proc.~SPIE}}   

\let\astap=\aap
\let\apjlett=\apjl
\let\apjsupp=\apjs
\let\applopt=\ao

{}

\end{document}